\documentclass[ALICE,manyauthors,nodoublepage]{cernphprep}

\usepackage[english]{babel}
\usepackage{graphicx}
\usepackage{verbatim}
\usepackage{amsfonts}
\usepackage{makeidx}
\usepackage{color}
\usepackage{amsmath}
\usepackage{lineno}
\usepackage{epstopdf}
\usepackage{hyperref}
\usepackage{cite}
\newcommand{\PbPb}{\mbox{Pb--Pb}}
\newcommand{\pPb}{\mbox{p--Pb}}

\newcommand{\sqrts}{\sqrt{s}}
\newcommand{\sqrtsNN}{\sqrt{s_{\rm \scriptscriptstyle NN}}}

\newcommand{\TeV}{\mathrm{TeV}}

\newcommand{\gev}{\mathrm{GeV}}
\newcommand{\tev}{\mathrm{TeV}}

\newcommand{\Raa}{R_{\rm AA}}
\newcommand{\RpPb}{R_{\rm pPb}}
\newcommand{\RAA}{R_{\rm AA}}

\newcommand{\pt}{p_{\rm T}}

\newcommand{\DtoKpi}{{\rm D^0\to K^-\pi^+}}
\newcommand{\DtoKpipi}{{\rm D^+\to K^-\pi^+\pi^+}}
\newcommand{\DstartoDpi}{{\rm D^{*+}\to D^0\pi^+}}
\newcommand{\DstophipitoKKpi}{{\rm D_s^{+}\to \phi\pi^+\to K^-K^+\pi^+}}

\newcommand{\Dzero}{{\rm D^0}}
\newcommand{\Dstar}{{\rm D^{*+}}}
\newcommand{\Dplus}{{\rm D^+}}
\newcommand{\Ds}{{\rm D_s^+}}

\begin{document}
\begin{titlepage}
\PHyear{2014}
\PHnumber{90}              
\PHdate{08 May}  
\title{Measurement of prompt D-meson production\\ in $\pPb$ collisions at 
$\mathbf{\sqrtsNN~=~5.02~\TeV}$}
\Collaboration{ALICE Collaboration
         \thanks{See Appendix~\ref{app:collab} for the list of collaboration 
                      members}}
\ShortAuthor{ALICE Collaboration}
\ShortTitle{D-meson production in $\pPb$ collisions at $\sqrtsNN~=~5.02~\TeV$}

\begin{abstract}
The $\pt$-differential production cross sections of the
prompt charmed mesons $\Dzero$, $\Dplus$, $\Dstar$ and $\Ds$ and their charge conjugate in the rapidity interval 
$-0.96<y_{\rm cms}<0.04$ were measured in $\pPb$ collisions at a 
centre-of-mass energy $\sqrtsNN=5.02~\tev$ with the ALICE detector at the LHC.
The nuclear modification factor $\RpPb$, quantifying the D-meson yield in p--Pb collisions relative to the yield in pp collisions scaled by the number of binary nucleon-nucleon collisions, is compatible within the 15-20\% uncertainties with unity in the transverse momentum 
interval $1<\pt<24~\gev/c$. No significant difference among the $\RpPb$ of the four D-meson species is observed. The results are described within uncertainties by theoretical calculations that include initial-state effects.
The measurement adds experimental evidence that the modification of the momentum spectrum of D mesons observed in Pb--Pb collisions with respect to pp collisions is due to strong final-state effects induced by hot partonic matter.
\end{abstract}
\end{titlepage}
In hadronic collisions heavy quarks are produced in scattering processes with large momentum transfer. 
Theoretical predictions based on perturbative Quantum Chromo-Dynamics (QCD) describe the $\pt$-differential charm production cross sections in pp collisions at different energies~\cite{GMVFNS, fonll, Maciula:2013wg}. 

The interpretation of heavy-ion collision experimental results is consistent with the formation of a high-density colour-deconfined medium, the Quark-Gluon Plasma (QGP)~\cite{Adams:2005dq,Aamodt:2010pa}. Heavy quarks are sensitive to the transport properties of the medium since they are produced on a short time-scale and traverse the medium interacting with its constituents. 
 In Pb--Pb collisions at $\sqrtsNN=2.76$~TeV, the D-meson nuclear modification factor $\RAA$, defined as ratio of the yield in nucleus-nucleus collisions to that observed in the pp ones scaled by the number of binary nucleon-nucleon collisions, indicates a strong suppression of the D-meson yield for $\pt \gtrsim 2~{\rm GeV}/c$~\cite{aliceDRAA}. The suppression is interpreted as due to in-medium energy loss~\cite{Uphoff:2012gb,whdg,vitev,He:2014cla}.
A complete understanding of the Pb--Pb results requires an understanding of cold-nuclear-matter effects in the initial and final state, which can be accessed by studying p--Pb collisions assuming that the QGP is not formed in these collisions. 
In the initial state, the nuclear environment affects the quark and gluon distributions, which are modified in bound nucleons depending on the  parton fractional momentum $x$ and the atomic mass number A~\cite{ARNEODO,Malace}. At LHC energies, the most relevant effect is gluon saturation at low $x$, which can modify the D-meson production significantly at low $\pt$. This effect can be described either by means of calculations based on phenomenological modification of the Parton Distribution Functions (PDFs) \cite{EPS09, deFlorianDS, HiraiHKN} or with the Colour Glass Condensate (GCG) effective theory~\cite{CGC2, Tribedy, Albacete, Rezaeian}. 
Partons can also lose energy in the initial stages of the collision via initial-state radiation, thus modifying the centre-of-mass energy of the partonic system \cite{Vitev:2007ve}, or experience transverse momentum broadening due to multiple soft collisions before the ${\rm c} \bar{\rm c}$ pair is produced \cite{Lev:1983hh, Wang:1998ww, Kopeliovich:2002yh}. 
Recent calculations of parton energy loss in the nuclear medium suggest that the formed ${\rm c} \bar{\rm c}$ pair is also affected by these processes in p--Pb collisions~\cite{Arleo}. The presence of final-state effects in small collision systems is suggested by recent studies on long-range correlations of charged hadrons~\cite{CMS:2012lrc, Abelev:2012ola, ABELEV:2013wsa, Aad:2012gla} in p--Pb collisions, by results on the species-dependent nuclear modification factors of pions, kaons and protons~\cite{Adler:2006xd} in d--Au collisions and on the larger suppression of the $\psi '$ meson with respect to the ${\rm J}/\psi$ in both d--Au \cite{Adare:2013ezl} and p--Pb \cite{Abelev:2014zpa} collisions.

Previous studies to address cold-nuclear-matter effects in heavy-flavour production were carried out at RHIC by measuring the production of leptons from heavy-flavour hadrons decays in d--Au collisions at $\sqrtsNN~=200$ GeV \cite{phenixdAu, phenixdAumuons, STARDmesons_dAu}. PHENIX measured an enhancement of about 40\% of the heavy-flavour decay electrons in the 20\% most central d--Au collisions with respect to pp collisions~\cite{phenixdAu}. A description of this result in terms of  hydrodynamic flow in small collision systems was recently proposed~\cite{AnneSickles}. PHENIX also measured an enhancement (suppression) of heavy-flavour decay muons at backward (forward) rapidities in d--Au collisions~\cite{phenixdAumuons}. The difference observed in the two rapidity regions exceeds predictions based on initial parton density modifications, suggesting the presence of other cold-nuclear-matter effects. 
The measurement of fully reconstructed charmed hadrons in p--Pb collisions at the LHC
can shed light on the different aspects of cold-nuclear-matter effects mentioned above and, in particular, can clarify whether the observed suppression of D-meson production in Pb--Pb collisions is a genuine hot QCD matter effect.

In this Letter, we present the measurement of the cross sections and of the nuclear modification factors, $\RpPb$, of prompt $\Dzero$, $\Dplus$, $\Dstar$ and $\Ds$ mesons in $\pPb$ collisions at $\sqrtsNN=5.02~\rm{TeV}$ performed with the ALICE detector~\cite{aliceJINST, alicePerformance} at the LHC.
D mesons were reconstructed in the rapidity interval $|y_{\rm lab}| < 0.5$ via their hadronic decay channels $\DtoKpi$ (with a branching ratio (BR) of $3.88 \pm 0.05\%$),
$\DtoKpipi$ (BR of $9.13 \pm 0.19\%$), $\DstartoDpi$ (BR of $67.7\pm
0.5\%$) and $\DstophipitoKKpi$ (BR of $2.28 \pm 0.12\%$)~\cite{PDG} and their charge conjugates.
Because of the different energies per nucleon of the proton and
the lead beams, the nucleon--nucleon centre-of-mass frame was moving with a rapidity $|\Delta y_{\rm  \scriptscriptstyle NN}| = 0.465$ in the proton beam direction (positive rapidities), leading to the rapidity coverage $-0.96<y_{\rm cms}<0.04$. 

Charged particles were reconstructed and identified with the central barrel detectors located within a 0.5~T solenoid magnet. 
Tracks were reconstructed with the Inner Tracking System (ITS) 
and the Time Projection Chamber (TPC). Particle IDentification (PID) was based on the specific energy
loss ${\rm d}E/{\rm d} x$ in the TPC gas and on the time of flight from the
interaction point to the Time Of Flight (TOF) detector. 
The analysis was performed by using $\pPb$ data collected in 2013 with a minimum-bias trigger that required  the arrival of bunches from both directions and coincident signals in both scintillator arrays of the V0 detector, covering the regions $2.8<\eta< 5.1$ and $-3.7<\eta<-1.7$.
Events were selected off-line by using the timing information from the V0 and the Zero Degree Calorimeters to remove background due to beam-gas interactions. Only events with a primary vertex
reconstructed within $\pm10~\rm{cm}$ from the centre of the
detector along the beam line were considered. About $10^8$ events, corresponding to an integrated luminosity of $(48.6 \pm 1.6)~\rm{\mu b^{-1}}$, passed the selection criteria.

D-meson selection was based on the reconstruction of decay vertices displaced from the interaction vertex, exploiting the separation of a few hundred micrometers typical of the D-meson weak decays, as described in ~\cite{aliceDpp276, aliceDpp7, aliceDRAA, aliceDs7}.
$\Dzero$, $\Dplus$ and $\Ds$ candidates were defined using pairs or
triplets of tracks with the proper charge sign combination. Tracks were required to have $|\eta| < 0.8$, $\pt
> 0.4~{\rm GeV}/c$, at least 70 out of 159 associated space points in the TPC and
at least 2 out of 6 hits in the ITS, out of which at least one in the two innermost layers. $\Dstar$ candidates were formed combining $\Dzero$ candidates with
tracks with $|\eta| < 0.8$, $\pt > 0.1~{\rm GeV}/c$ and at least three
associated hits in the ITS.  
The selection strategy was based on the displacement of the tracks from the interaction vertex and the pointing of the reconstructed D-meson momentum to the primary vertex. At low-$\pt$, further background rejection was obtained by identifying charged kaons with the TPC and TOF by applying cuts in units of resolution ($\pm 3 \sigma$) around the expected mean values of ${\rm d}E/{\rm d} x$ and time of flight.
For $\Ds$ candidate selection, the invariant mass of at least one of the two
opposite-charge track pairs was required to be compatible with the mass of 
the $\phi$ meson ($\pm 2 \sigma$).

The total cross section for hard processes $\sigma_{\rm pA}^{\rm hard}$ in proton--nucleus collisions scales as $\sigma_{\rm pA}^{\rm hard} = {\rm A}~\sigma_{\rm NN}^{\rm hard}$~\cite{d'Enterria:2003qs}, where  
$\sigma_{\rm NN}^{\rm hard}$ is the equivalent cross section in pp collisions. Therefore, the $\RpPb$ for prompt D mesons is given by
\begin{equation}
R_{{\rm pPb}}=\frac{\left(\frac{{\rm d}\sigma}{{\rm d}\pt}\right)_{{\rm
    pPb}}}{{\rm A} \times\left(\frac{{\rm d}\sigma}{{\rm d}\pt}\right)_{{\rm pp}}}.
\end{equation} 
The production cross sections of prompt D mesons (not coming from beauty meson decays) were obtained as (e.g. for $\Dplus$)
\begin{equation}
\left.\frac{\rm{d}\sigma^{\rm{D^+}}}{{\rm    d}\pt}\right|_{\scriptscriptstyle |y_{\rm lab}|<0.5}=\frac{f_{\rm{prompt}}\cdot N^{\rm
    D^{\pm}}_{\rm raw}\mid_{|y_{\rm lab}|<y_{\rm \scriptscriptstyle fid}}}{2 \alpha_{\rm \scriptscriptstyle y}  \Delta\pt ({\rm Acc} \times
   \epsilon)_{{\rm \scriptscriptstyle prompt}}\cdot {\rm{BR}}\cdot L_{\rm \scriptscriptstyle int}}.
\end{equation}
$N^{{\rm D^{\pm}}}_{\rm raw}$ is the raw yield extracted in a given $\pt$ interval (of width
$\Delta\pt$) by means of a fit to the invariant mass distribution of the D-meson candidates. $f_{{\rm prompt}}$ is the prompt fraction of the
raw yield. $({\rm Acc}\times\epsilon)_{{\rm prompt}}$ is the
geometrical acceptance multiplied by the reconstruction and selection efficiency of prompt D mesons.
The factor $\alpha_{\rm \scriptscriptstyle y} = y_{\rm fid}/0.5$ normalizes the yields, measured in $|y_{\rm lab}|<y_{\rm fid}$, to one unit of rapidity $|y_{\rm lab}|<0.5$. $y_{{\rm fid}}$ is the $\pt$-dependent fiducial acceptance cut ($y_{{\rm fid}}$ increases from 0.5 at $\pt=0$ to 0.8 at $\pt=5~{\rm GeV}/c$ and becomes constant at 0.8 for $\pt>5~{\rm GeV}/c$). The cross sections are given for particles;
thus, a factor $1/2$ was added to take into account that both
particles and anti-particles are counted in the raw yield. The integrated luminosity, $L_{{\rm int}}$, was computed as $N_{{\rm
    pPb,MB}}/\sigma_{{\rm pPb, MB}}$ where $N_{{\rm pPb,MB}}$ is the number of p--Pb collisions passing the minimum-bias trigger condition and $\sigma_{{\rm pPb,MB}}$ is the cross section of the V0
trigger, which was measured to be $2.09~{\rm b} \pm 3.5\%$ (syst) with the p--Pb van der Meer scan~\cite{vdm}. The minimum-bias trigger is 100\% efficient for D mesons with $\pt> 1~{\rm GeV}/c$ and $|y_{\rm lab}| < 0.5$.

The acceptance-times-efficiency $({\rm Acc}\times\epsilon)$ corrections were determined by using a
Monte Carlo simulation. Proton--lead collisions were produced
using the HIJING v1.36~\cite{hijing} event generator. 
A ${\rm c}\bar{\rm c}$ or ${\rm b} \bar{\rm b}$ pair was added in each event using the PYTHIA v6.4.21~\cite{pythia} generator with Perugia-0 tuning~\cite{Skands:2010ak}. 
The generated particles were transported through the ALICE detector using GEANT3 \cite{geant}. 
The efficiency for D-meson reconstruction and selection varies from 0.5-1\% for $\pt < 2~{\rm GeV}/c$ to 20-30\% for $\pt > 12~{\rm GeV}/c$ because of the larger displacement of the decay vertex of high-$\pt$ candidates due to the Lorentz boost. Hence the generated D-meson spectrum used to calculate the efficiencies was tuned to reproduce the shape given by Fixed-Order Next-To-Leading-Log resummation (FONLL)~\cite{fonll} calculations at $\sqrts=5.02$~TeV in each $\pt$ interval. The efficiency depends also on the multiplicity of charged particles produced in the collision since the primary vertex resolution, and consequently the resolution of the topological selection variables, improves with increasing multiplicity. This dependence is different for each meson species and  $\pt$ interval: e.g. the $\Dzero$ efficiency in $5 < \pt < 8~{\rm GeV}/c$ increases by a factor 1.5 for low multiplicity events until it becomes constant at about 20 reconstructed primary particles.
Therefore, the efficiency was calculated by weighting the simulated events according to their charged particle multiplicity in order to reproduce the multiplicity distribution observed in data.\\
The fraction of prompt D mesons, $f_{\rm prompt}$, was estimated as in Ref.~\cite{aliceDRAA} by using the beauty production cross section from FONLL calculations \cite{fonll}, the ${\rm B} \rightarrow {\rm D } + X$ decay kinematics from the EvtGen package \cite{evtgen} and the reconstruction and selection efficiency for D mesons from B hadron decays. The $\RpPb$ of prompt and feed-down D mesons were assumed to be equal and were varied in the range $0.9 < R_{\rm pPb}^{\rm feed-down} / R_{\rm pPb}^{\rm prompt}<1.3$ to evaluate the systematic uncertainties.
This range was chosen considering the predictions from calculations including initial state effects based on the Eskola-Paukkunen-Salgado 2009 (EPS09)~\cite{EPS09} parameterizations  of the nuclear modification of the PDF and CGC~\cite{CGC2}.

The reference pp cross sections at $\sqrt{s}=5.02~{\rm TeV}$ were obtained by a perturbative-QCD-based energy scaling of the $\pt$-differential cross sections measured at $\sqrt{s}=7~\rm{TeV}$~\cite{aliceDpp7}.
The scaling factor for each D-meson species was determined as the ratio of the 
cross sections from the FONLL calculations at 5.02 and 7 TeV. The uncertainty on the scaling factor was evaluated by varying the calculation parameters as described in Ref.~\cite{scaling} and it ranges from $^{+17.5\%}_{-4\%}$ at $\pt=1~{\rm GeV}/c$ to about $\pm 3\%$ for $\pt>8~{\rm GeV}/c$. In addition, the pp reference is affected by the uncertainty coming from the 7 TeV measurement ($\sim$17\%)~\cite{aliceDpp7}.
Since the $\Dzero$ cross section in pp collisions in the $1 < \pt < 2~{\rm GeV}/c$ interval was measured at both 7 and 2.76 TeV, both results were scaled to 5.02 TeV and averaged considering their relative statistical and systematic uncertainties as weights. 
Since the current measurement of the ALICE $\Dzero$ pp cross section at $\sqrt{s}=7~{\rm TeV}$ is limited to $\pt = 16~{\rm GeV}/c$, the cross section was extrapolated to higher $\pt$ using the spectrum predicted by FONLL~\cite{fonll} scaled to match pp data in $5 < \pt < 16~{\rm GeV}/c$. Then the $\Dzero$ cross section at 7 TeV in $16 < \pt < 24~{\rm GeV}/c$ was scaled to 5.02 TeV.

The systematic uncertainties on the D-meson cross sections include contributions from yield extraction (from 2\% to 17\% depending on $\pt$ and D-meson species), an imperfect description of the cut variables in the simulation (from 5\% to 8\% for $\Dzero$, $\Dplus$ and $\Dstar$, $\sim$20\% for $\Ds$), tracking efficiency (3\% for each track), simulated $\pt$ shapes (from 2\% to 3\% depending on $\pt$ and D-meson species), and the subtraction of feed-down D mesons from B decays (from 4\% and 40\% depending on $\pt$ and D-meson species). For the $\Dzero$ meson, the yield extraction systematic uncertainty also includes the contribution to the raw yield of signal candidates reconstructed assigning the wrong mass to the final-state hadrons. This contribution, which is strongly reduced by the PID selection, was estimated to be 3\%(4\%) at low(high) $\pt$ based on the invariant mass distribution of these candidates in the simulation.
Details of the procedure for the systematic uncertainty estimation are reported in Refs.~\cite{aliceDpp276, aliceDpp7, aliceDRAA, aliceDs7}.
The measured cross sections have a global systematic uncertainty due to the determination of the integrated luminosity (3.7\%~\cite{vdm}) and to the branching ratio~\cite{PDG}.
For the $\RpPb$, the pp and p--Pb uncertainties were added in quadrature except for the branching ratio uncertainty, which cancels out in the ratio, and the feed-down contribution, which partially cancels out.

\begin{figure}[!t]
\begin{center}
\includegraphics[width=.5\textwidth]{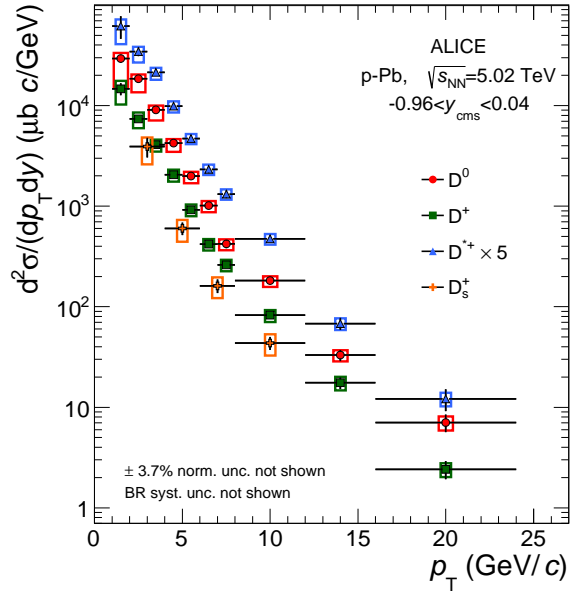}
\caption{$\pt$-differential inclusive production cross section of prompt
$\Dzero$, $\Dplus$, $\Dstar$ and $\Ds$ mesons in $\pPb$ collisions
at $\sqrtsNN=5.02~\tev$. Statistical uncertainties (bars) and systematic uncertainties (boxes) are shown.}
\label{fig:crosssec} 
\end{center}
\end{figure}
\begin{figure*}[!hbt]
\begin{center}
\includegraphics[width=1\textwidth]{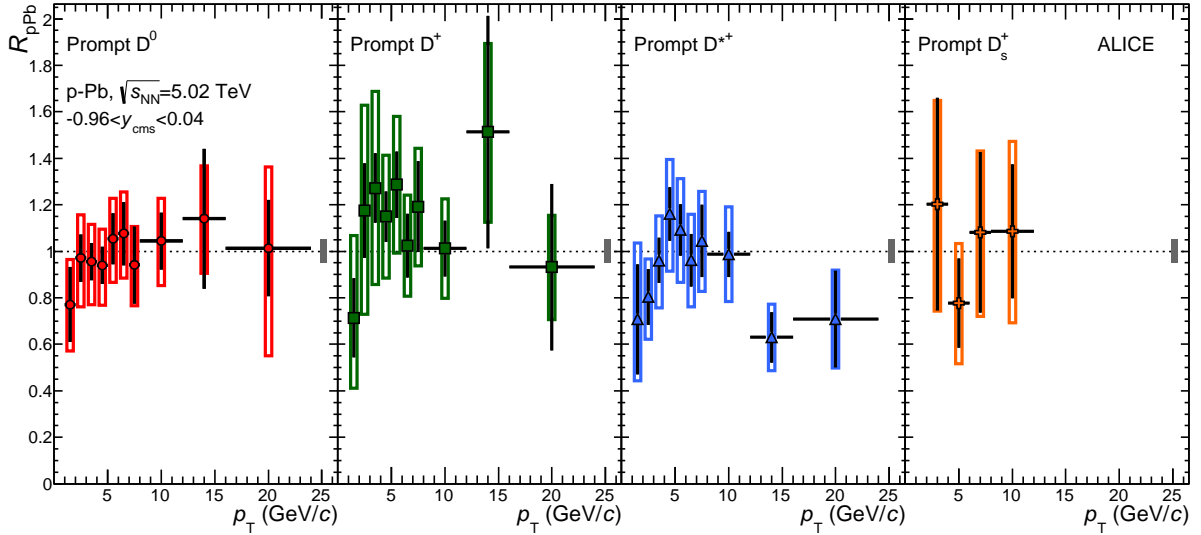}
\caption{$\RpPb$ as a function of $\pt$ for prompt  $\Dzero$, $\Dplus$, $\Dstar$ and 
$\Ds$ mesons in $\pPb$ collisions at $\sqrtsNN=5.02~\tev$. Statistical (bars), systematic (empty boxes), and normalization (full box) uncertainties are shown.}
\label{fig:rppb4} 
\end{center}
\end{figure*}
\begin{figure}[!htb]
\begin{center}
\includegraphics[width=.5\textwidth]{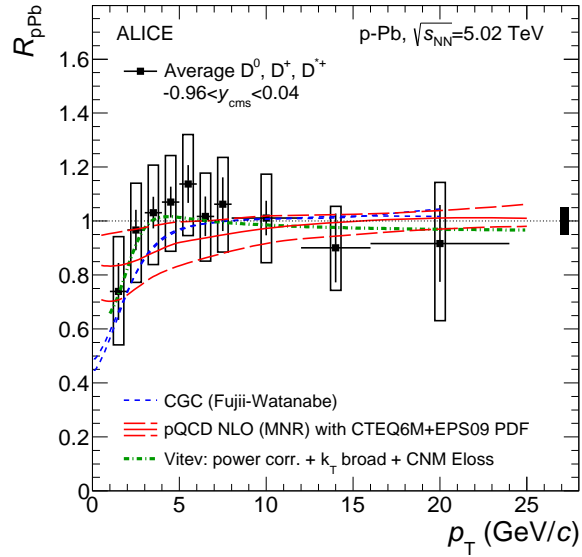} 
\caption{Average $\RpPb$ of prompt $\Dzero$, $\Dplus$ 
and $\Dstar$ mesons as a function of $\pt$ compared to model calculations. Statistical (bars), systematic (empty boxes), and normalization (full box) uncertainties are shown.}
\label{fig:rppbMod}
\end{center}
\end{figure}
\begin{figure}[!htb]
\begin{center}
\includegraphics[width=.5\textwidth]{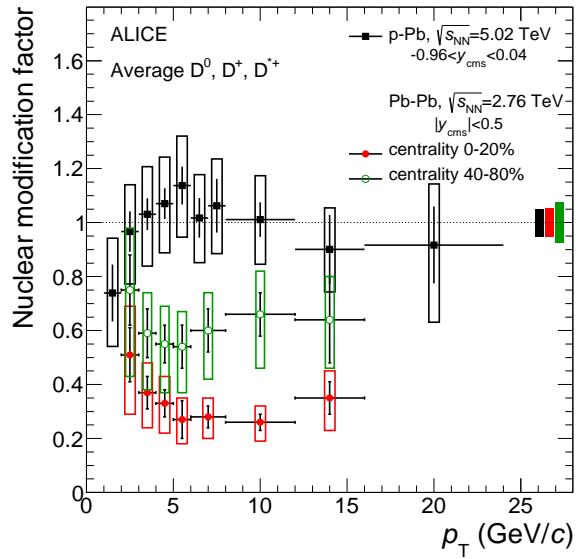} 
\caption{Average $\RpPb$ of prompt $\Dzero$, $\Dplus$ 
and $\Dstar$ mesons as a function of $\pt$ compared to D-meson $\Raa$ in the 20\% most central and in the 40\%-80\% Pb--Pb collisions at $\sqrtsNN=2.76~\tev$ from Ref.~\cite{aliceDRAA}. Statistical (bars), systematic (empty boxes), and normalization (full boxes) uncertainties are shown.}
\label{fig:rppbrpbpb}
\end{center}
\end{figure}
The $\pt$-differential production cross sections of prompt $\Dzero$, $\Dplus$, $\Dstar$ and $\Ds$ mesons are shown in Fig.~\ref{fig:crosssec}. The relative abundances of D mesons in p--Pb collisions are compatible within uncertainties with those measured in pp, $e$p and $e^{+}e^{-}$ collisions at different energies~\cite{aliceDs7}. 
The $\RpPb$ of the four D-meson species, shown in Fig.~\ref{fig:rppb4}, are consistent, and they are compatible with unity within the uncertainties in the measured $p_{\rm T}$ range. D-meson production in p--Pb collisions is consistent within statistical and systematic uncertainties with the binary collision scaling of the production in pp collisions. Moreover, within the uncertainties, the $\Ds$ nuclear modification factor is compatible with that of non-strange D mesons. The average of the $\RpPb$ of $\Dzero$, $\Dplus$ and $\Dstar$ in the $p_{\rm T}$ range $1<p_{\rm T} < 24~{\rm GeV}/c$ was calculated using the relative statistical uncertainties as weights. The systematic error on the average was calculated by propagating the uncertainties through the weighted average, where the contributions from tracking efficiency, B feed-down correction, and scaling of the pp reference were taken as fully correlated among the three species. 
Figure~\ref{fig:rppbMod} shows the average $\RpPb$ compared to theoretical calculations. 
Predictions based either on next-to-leading order (NLO) pQCD calculations (MNR \cite{MNR}) of D-meson production, including the EPS09~\cite{EPS09} nuclear modification of the CTEQ6M PDF~\cite{CTEQ6M}, or on calculations based on the Color Glass Condensate~\cite{CGC2} can describe the measurement by considering only initial-state effects. Data are also well described by calculations which include cold-nuclear-matter energy loss, nuclear shadowing and $k_{\rm T}$-broadening~\cite{vitev}.
The possible effects due to the formation of a hydrodynamically expanding medium as calculated in Ref.~\cite{AnneSickles} are expected to be small in minimum-bias collisions at LHC energies. The present uncertainties of the measurement do not allow any sensitivity to this effect.
In Fig.~\ref{fig:rppbrpbpb}, the average $\Raa$ of prompt D mesons in central (0-20\%) and in semi-peripheral (40\%-80\%) Pb--Pb collisions at $\sqrtsNN=2.76~{\rm TeV}$ \cite{aliceDRAA} is reported along with the average $\RpPb$ of prompt D mesons in p--Pb collisions at $\sqrtsNN=5.02~{\rm TeV}$, showing that cold-nuclear-matter effects are smaller than the uncertainties for $\pt \gtrsim 3~{\rm GeV}/c$. In addition, as reported in Ref.~\cite{aliceDRAA}, the same EPS09 nuclear PDF parametrization that describes the D-meson $\RpPb$ results predicts small initial state effects (less than 10\% for $\pt >5~{\rm GeV}/c$) for $\PbPb$ collisions.
As a consequence, the suppression observed in central Pb--Pb collisions for $\pt \gtrsim 2~{\rm GeV}/c$ is predominantly induced by final-state effects, e.g. the charm energy loss in the medium~\cite{Uphoff:2012gb,whdg,vitev,He:2014cla}.

In summary, we reported the measurement of the D-meson cross section and nuclear modification factor in p--Pb collisions at $\sqrtsNN = 5.02$~TeV. The latter is consistent within uncertainties of about 15\%-20\% with unity and is compatible with theoretical calculations including gluon saturation. Thus, the suppression of D mesons with $\pt \gtrsim 2~{\rm GeV}/c$ observed in Pb--Pb collisions cannot be explained in terms of initial-state effects but is
due to strong final-state effects induced by hot partonic matter.  

%\newpage
\newenvironment{acknowledgement}{\relax}{\relax}
\begin{acknowledgement}
\section*{Acknowledgements}
The ALICE Collaboration would like to thank all its engineers and technicians for their invaluable contributions to the construction of the experiment and the CERN accelerator teams for the outstanding performance of the LHC complex.
%\\
The ALICE Collaboration would like to thank M.~Cacciari for 
providing the pQCD predictions used for the feed-down correction and the energy 
scaling, and I.~Vitev, H.~Fujii and K.~Watanabe
for making available their predictions for the nuclear modification factor.
%\\
The ALICE Collaboration gratefully acknowledges the resources and support provided by all Grid centres and the Worldwide LHC Computing Grid (WLCG) collaboration.
%\\
The ALICE Collaboration acknowledges the following funding agencies for their support in building and
running the ALICE detector:
 %\\
State Committee of Science,  World Federation of Scientists (WFS)
and Swiss Fonds Kidagan, Armenia,
 %\\
Conselho Nacional de Desenvolvimento Cient\'{\i}fico e Tecnol\'{o}gico (CNPq), Financiadora de Estudos e Projetos (FINEP),
Funda\c{c}\~{a}o de Amparo \`{a} Pesquisa do Estado de S\~{a}o Paulo (FAPESP);
 %\\
National Natural Science Foundation of China (NSFC), the Chinese Ministry of Education (CMOE)
and the Ministry of Science and Technology of China (MSTC);
 %\\
Ministry of Education and Youth of the Czech Republic;
 %\\
Danish Natural Science Research Council, the Carlsberg Foundation and the Danish National Research Foundation;
 %\\
The European Research Council under the European Community's Seventh Framework Programme;
 %\\
Helsinki Institute of Physics and the Academy of Finland;
 %\\
French CNRS-IN2P3, the `Region Pays de Loire', `Region Alsace', `Region Auvergne' and CEA, France;
 %\\
German BMBF and the Helmholtz Association;
%\\
General Secretariat for Research and Technology, Ministry of
Development, Greece;
%\\
Hungarian OTKA and National Office for Research and Technology (NKTH);
 %\\
Department of Atomic Energy and Department of Science and Technology of the Government of India;
 %\\
Istituto Nazionale di Fisica Nucleare (INFN) and Centro Fermi -
Museo Storico della Fisica e Centro Studi e Ricerche "Enrico
Fermi", Italy;
 %\\
MEXT Grant-in-Aid for Specially Promoted Research, Ja\-pan;
 %\\
Joint Institute for Nuclear Research, Dubna;
 %\\
%Korea Foundation for International Cooperation of Science and Technology (KICOS);
National Research Foundation of Korea (NRF);
 %\\
CONACYT, DGAPA, M\'{e}xico, ALFA-EC and the EPLANET Program
(European Particle Physics Latin American Network)
 %\\
Stichting voor Fundamenteel Onderzoek der Materie (FOM) and the Nederlandse Organisatie voor Wetenschappelijk Onderzoek (NWO), Netherlands;
 %\\
Research Council of Norway (NFR);
 %\\
Polish Ministry of Science and Higher Education;
 %\\
National Science Centre, Poland;
 %\\
 Ministry of National Education/Institute for Atomic Physics and CNCS-UEFISCDI - Romania;
 %\\
Ministry of Education and Science of Russian Federation, Russian
Academy of Sciences, Russian Federal Agency of Atomic Energy,
Russian Federal Agency for Science and Innovations and The Russian
Foundation for Basic Research;
 %\\
Ministry of Education of Slovakia;
 %\\
Department of Science and Technology, South Africa;
 %\\
CIEMAT, EELA, Ministerio de Econom\'{i}a y Competitividad (MINECO) of Spain, Xunta de Galicia (Conseller\'{\i}a de Educaci\'{o}n),
CEA\-DEN, Cubaenerg\'{\i}a, Cuba, and IAEA (International Atomic Energy Agency);
 %\\
Swedish Research Council (VR) and Knut $\&$ Alice Wallenberg
Foundation (KAW);
 %\\
Ukraine Ministry of Education and Science;
 %\\
United Kingdom Science and Technology Facilities Council (STFC);
 %\\
The United States Department of Energy, the United States National
Science Foundation, the State of Texas, and the State of Ohio.

\end{acknowledgement}

\newpage
\appendix
\section{The ALICE Collaboration}
\label{app:collab}

% Collaboration: CERN-LHC-ALICE
% Generation Date is 2014/Apr/29

% How to use:
%%%%%%%%% appendix with author list
%\appendix
%\section{The ALICE Collaboration}
%\label{app:collab}
%\input{authors-list.tex}  %%%%%%% get the latest version before submitting

\begingroup
\small
\begin{flushleft}
B.~Abelev\Irefn{org71}\And
J.~Adam\Irefn{org37}\And
D.~Adamov\'{a}\Irefn{org79}\And
M.M.~Aggarwal\Irefn{org83}\And
G.~Aglieri~Rinella\Irefn{org34}\And
M.~Agnello\Irefn{org107}\textsuperscript{,}\Irefn{org90}\And
A.~Agostinelli\Irefn{org26}\And
N.~Agrawal\Irefn{org44}\And
Z.~Ahammed\Irefn{org126}\And
N.~Ahmad\Irefn{org18}
% ...
\renewcommand\labelenumi{\textsuperscript{\theenumi}~}

\section*{Affiliation notes}
\renewcommand\theenumi{\roman{enumi}}
\begin{Authlist}
\item \Adef{0}Deceased
\item \Adef{idp29610640}{Also at: St. Petersburg State Polytechnical University}
\item \Adef{idp31513984}{Also at: Department of Applied Physics, Aligarh Muslim University, Aligarh, India}
\item \Adef{idp32200928}{Also at: M.V. Lomonosov Moscow State University, D.V. Skobeltsyn Institute of Nuclear Physics, Moscow, Russia}
\item \Adef{idp32450016}{Also at: University of Belgrade, Faculty of Physics and "Vin\v{c}a" Institute of Nuclear Sciences, Belgrade, Serbia}
\item \Adef{idp32761744}{Permanent Address: Permanent Address: Konkuk University, Seoul, Korea}
\item \Adef{idp33313056}{Also at: Institute of Theoretical Physics, University of Wroclaw, Wroclaw, Poland}
\item \Adef{idp34232864}{Also at: University of Kansas, Lawrence, KS, United States}
\end{Authlist}

\section*{Collaboration Institutes}
\renewcommand\theenumi{\arabic{enumi}~}
\begin{Authlist}

\item \Idef{org1}A.I. Alikhanyan National Science Laboratory (Yerevan Physics Institute) Foundation, Yerevan, Armenia
\item \Idef{org2}Benem\'{e}rita Universidad Aut\'{o}noma de Puebla, Puebla, Mexico
\item \Idef{org3}Bogolyubov Institute for Theoretical Physics, Kiev, Ukraine
\item \Idef{org4}Bose Institute, Department of Physics and Centre for Astroparticle Physics and Space Science (CAPSS), Kolkata, India
\item \Idef{org5}Budker Institute for Nuclear Physics, Novosibirsk, Russia
\item \Idef{org6}California Polytechnic State University, San Luis Obispo, CA, United States
\item \Idef{org7}Central China Normal University, Wuhan, China
\item \Idef{org8}Centre de Calcul de l'IN2P3, Villeurbanne, France
\item \Idef{org9}Centro de Aplicaciones Tecnol\'{o}gicas y Desarrollo Nuclear (CEADEN), Havana, Cuba
\item \Idef{org10}Centro de Investigaciones Energ\'{e}ticas Medioambientales y Tecnol\'{o}gicas (CIEMAT), Madrid, Spain
\item \Idef{org11}Centro de Investigaci\'{o}n y de Estudios Avanzados (CINVESTAV), Mexico City and M\'{e}rida, Mexico
\item \Idef{org12}Centro Fermi - Museo Storico della Fisica e Centro Studi e Ricerche ``Enrico Fermi'', Rome, Italy
\item \Idef{org13}Chicago State University, Chicago, USA
\item \Idef{org14}Commissariat \`{a} l'Energie Atomique, IRFU, Saclay, France
\item \Idef{org15}COMSATS Institute of Information Technology (CIIT), Islamabad, Pakistan
\item \Idef{org16}Departamento de F\'{\i}sica de Part\'{\i}culas and IGFAE, Universidad de Santiago de Compostela, Santiago de Compostela, Spain
\item \Idef{org17}Department of Physics and Technology, University of Bergen, Bergen, Norway
\item \Idef{org18}Department of Physics, Aligarh Muslim University, Aligarh, India
\item \Idef{org19}Department of Physics, Ohio State University, Columbus, OH, United States
\item \Idef{org20}Department of Physics, Sejong University, Seoul, South Korea
\item \Idef{org21}Department of Physics, University of Oslo, Oslo, Norway
\item \Idef{org22}Dipartimento di Fisica dell'Universit\`{a} 'La Sapienza' and Sezione INFN Rome, Italy
\item \Idef{org23}Dipartimento di Fisica dell'Universit\`{a} and Sezione INFN, Cagliari, Italy
\item \Idef{org24}Dipartimento di Fisica dell'Universit\`{a} and Sezione INFN, Trieste, Italy
\item \Idef{org25}Dipartimento di Fisica dell'Universit\`{a} and Sezione INFN, Turin, Italy
\item \Idef{org26}Dipartimento di Fisica e Astronomia dell'Universit\`{a} and Sezione INFN, Bologna, Italy
\item \Idef{org27}Dipartimento di Fisica e Astronomia dell'Universit\`{a} and Sezione INFN, Catania, Italy
\item \Idef{org28}Dipartimento di Fisica e Astronomia dell'Universit\`{a} and Sezione INFN, Padova, Italy
\item \Idef{org29}Dipartimento di Fisica `E.R.~Caianiello' dell'Universit\`{a} and Gruppo Collegato INFN, Salerno, Italy
\item \Idef{org30}Dipartimento di Scienze e Innovazione Tecnologica dell'Universit\`{a} del  Piemonte Orientale and Gruppo Collegato INFN, Alessandria, Italy
\item \Idef{org31}Dipartimento Interateneo di Fisica `M.~Merlin' and Sezione INFN, Bari, Italy
\item \Idef{org32}Division of Experimental High Energy Physics, University of Lund, Lund, Sweden
\item \Idef{org33}Eberhard Karls Universit\"{a}t T\"{u}bingen, T\"{u}bingen, Germany
\item \Idef{org34}European Organization for Nuclear Research (CERN), Geneva, Switzerland
\item \Idef{org35}Faculty of Engineering, Bergen University College, Bergen, Norway
\item \Idef{org36}Faculty of Mathematics, Physics and Informatics, Comenius University, Bratislava, Slovakia
\item \Idef{org37}Faculty of Nuclear Sciences and Physical Engineering, Czech Technical University in Prague, Prague, Czech Republic
\item \Idef{org38}Faculty of Science, P.J.~\v{S}af\'{a}rik University, Ko\v{s}ice, Slovakia
\item \Idef{org39}Frankfurt Institute for Advanced Studies, Johann Wolfgang Goethe-Universit\"{a}t Frankfurt, Frankfurt, Germany
\item \Idef{org40}Gangneung-Wonju National University, Gangneung, South Korea
\item \Idef{org41}Gauhati University, Department of Physics, Guwahati, India
\item \Idef{org42}Helsinki Institute of Physics (HIP), Helsinki, Finland
\item \Idef{org43}Hiroshima University, Hiroshima, Japan
\item \Idef{org44}Indian Institute of Technology Bombay (IIT), Mumbai, India
\item \Idef{org45}Indian Institute of Technology Indore, Indore (IITI), India
\item \Idef{org46}Inha University, Incheon, South Korea
\item \Idef{org47}Institut de Physique Nucl\'eaire d'Orsay (IPNO), Universit\'e Paris-Sud, CNRS-IN2P3, Orsay, France
\item \Idef{org48}Institut f\"{u}r Informatik, Johann Wolfgang Goethe-Universit\"{a}t Frankfurt, Frankfurt, Germany
\item \Idef{org49}Institut f\"{u}r Kernphysik, Johann Wolfgang Goethe-Universit\"{a}t Frankfurt, Frankfurt, Germany
\item \Idef{org50}Institut f\"{u}r Kernphysik, Westf\"{a}lische Wilhelms-Universit\"{a}t M\"{u}nster, M\"{u}nster, Germany
\item \Idef{org51}Institut Pluridisciplinaire Hubert Curien (IPHC), Universit\'{e} de Strasbourg, CNRS-IN2P3, Strasbourg, France
\item \Idef{org52}Institute for Nuclear Research, Academy of Sciences, Moscow, Russia
\item \Idef{org53}Institute for Subatomic Physics of Utrecht University, Utrecht, Netherlands
\item \Idef{org54}Institute for Theoretical and Experimental Physics, Moscow, Russia
\item \Idef{org55}Institute of Experimental Physics, Slovak Academy of Sciences, Ko\v{s}ice, Slovakia
\item \Idef{org56}Institute of Physics, Academy of Sciences of the Czech Republic, Prague, Czech Republic
\item \Idef{org57}Institute of Physics, Bhubaneswar, India
\item \Idef{org58}Institute of Space Science (ISS), Bucharest, Romania
\item \Idef{org59}Instituto de Ciencias Nucleares, Universidad Nacional Aut\'{o}noma de M\'{e}xico, Mexico City, Mexico
\item \Idef{org60}Instituto de F\'{\i}sica, Universidad Nacional Aut\'{o}noma de M\'{e}xico, Mexico City, Mexico
\item \Idef{org61}iThemba LABS, National Research Foundation, Somerset West, South Africa
\item \Idef{org62}Joint Institute for Nuclear Research (JINR), Dubna, Russia
\item \Idef{org63}Konkuk University, Seoul, South Korea
\item \Idef{org64}Korea Institute of Science and Technology Information, Daejeon, South Korea
\item \Idef{org65}KTO Karatay University, Konya, Turkey
\item \Idef{org66}Laboratoire de Physique Corpusculaire (LPC), Clermont Universit\'{e}, Universit\'{e} Blaise Pascal, CNRS--IN2P3, Clermont-Ferrand, France
\item \Idef{org67}Laboratoire de Physique Subatomique et de Cosmologie, Universit\'{e} Grenoble-Alpes, CNRS-IN2P3, Grenoble, France
\item \Idef{org68}Laboratori Nazionali di Frascati, INFN, Frascati, Italy
\item \Idef{org69}Laboratori Nazionali di Legnaro, INFN, Legnaro, Italy
\item \Idef{org70}Lawrence Berkeley National Laboratory, Berkeley, CA, United States
\item \Idef{org71}Lawrence Livermore National Laboratory, Livermore, CA, United States
\item \Idef{org72}Moscow Engineering Physics Institute, Moscow, Russia
\item \Idef{org73}National Centre for Nuclear Studies, Warsaw, Poland
\item \Idef{org74}National Institute for Physics and Nuclear Engineering, Bucharest, Romania
\item \Idef{org75}National Institute of Science Education and Research, Bhubaneswar, India
\item \Idef{org76}Niels Bohr Institute, University of Copenhagen, Copenhagen, Denmark
\item \Idef{org77}Nikhef, National Institute for Subatomic Physics, Amsterdam, Netherlands
\item \Idef{org78}Nuclear Physics Group, STFC Daresbury Laboratory, Daresbury, United Kingdom
\item \Idef{org79}Nuclear Physics Institute, Academy of Sciences of the Czech Republic, \v{R}e\v{z} u Prahy, Czech Republic
\item \Idef{org80}Oak Ridge National Laboratory, Oak Ridge, TN, United States
\item \Idef{org81}Petersburg Nuclear Physics Institute, Gatchina, Russia
\item \Idef{org82}Physics Department, Creighton University, Omaha, NE, United States
\item \Idef{org83}Physics Department, Panjab University, Chandigarh, India
\item \Idef{org84}Physics Department, University of Athens, Athens, Greece
\item \Idef{org85}Physics Department, University of Cape Town, Cape Town, South Africa
\item \Idef{org86}Physics Department, University of Jammu, Jammu, India
\item \Idef{org87}Physics Department, University of Rajasthan, Jaipur, India
\item \Idef{org88}Physik Department, Technische Universit\"{a}t M\"{u}nchen, Munich, Germany
\item \Idef{org89}Physikalisches Institut, Ruprecht-Karls-Universit\"{a}t Heidelberg, Heidelberg, Germany
\item \Idef{org90}Politecnico di Torino, Turin, Italy
\item \Idef{org91}Purdue University, West Lafayette, IN, United States
\item \Idef{org92}Pusan National University, Pusan, South Korea
\item \Idef{org93}Research Division and ExtreMe Matter Institute EMMI, GSI Helmholtzzentrum f\"ur Schwerionenforschung, Darmstadt, Germany
\item \Idef{org94}Rudjer Bo\v{s}kovi\'{c} Institute, Zagreb, Croatia
\item \Idef{org95}Russian Federal Nuclear Center (VNIIEF), Sarov, Russia
\item \Idef{org96}Russian Research Centre Kurchatov Institute, Moscow, Russia
\item \Idef{org97}Saha Institute of Nuclear Physics, Kolkata, India
\item \Idef{org98}School of Physics and Astronomy, University of Birmingham, Birmingham, United Kingdom
\item \Idef{org99}Secci\'{o}n F\'{\i}sica, Departamento de Ciencias, Pontificia Universidad Cat\'{o}lica del Per\'{u}, Lima, Peru
\item \Idef{org100}Sezione INFN, Bari, Italy
\item \Idef{org101}Sezione INFN, Bologna, Italy
\item \Idef{org102}Sezione INFN, Cagliari, Italy
\item \Idef{org103}Sezione INFN, Catania, Italy
\item \Idef{org104}Sezione INFN, Padova, Italy
\item \Idef{org105}Sezione INFN, Rome, Italy
\item \Idef{org106}Sezione INFN, Trieste, Italy
\item \Idef{org107}Sezione INFN, Turin, Italy
\item \Idef{org108}SSC IHEP of NRC Kurchatov institute, Protvino, Russia
\item \Idef{org109}SUBATECH, Ecole des Mines de Nantes, Universit\'{e} de Nantes, CNRS-IN2P3, Nantes, France
\item \Idef{org110}Suranaree University of Technology, Nakhon Ratchasima, Thailand
\item \Idef{org111}Technical University of Split FESB, Split, Croatia
\item \Idef{org112}The Henryk Niewodniczanski Institute of Nuclear Physics, Polish Academy of Sciences, Cracow, Poland
\item \Idef{org113}The University of Texas at Austin, Physics Department, Austin, TX, USA
\item \Idef{org114}Universidad Aut\'{o}noma de Sinaloa, Culiac\'{a}n, Mexico
\item \Idef{org115}Universidade de S\~{a}o Paulo (USP), S\~{a}o Paulo, Brazil
\item \Idef{org116}Universidade Estadual de Campinas (UNICAMP), Campinas, Brazil
\item \Idef{org117}University of Houston, Houston, TX, United States
\item \Idef{org118}University of Jyv\"{a}skyl\"{a}, Jyv\"{a}skyl\"{a}, Finland
\item \Idef{org119}University of Liverpool, Liverpool, United Kingdom
\item \Idef{org120}University of Tennessee, Knoxville, TN, United States
\item \Idef{org121}University of Tokyo, Tokyo, Japan
\item \Idef{org122}University of Tsukuba, Tsukuba, Japan
\item \Idef{org123}University of Zagreb, Zagreb, Croatia
\item \Idef{org124}Universit\'{e} de Lyon, Universit\'{e} Lyon 1, CNRS/IN2P3, IPN-Lyon, Villeurbanne, France
\item \Idef{org125}V.~Fock Institute for Physics, St. Petersburg State University, St. Petersburg, Russia
\item \Idef{org126}Variable Energy Cyclotron Centre, Kolkata, India
\item \Idef{org127}Vestfold University College, Tonsberg, Norway
\item \Idef{org128}Warsaw University of Technology, Warsaw, Poland
\item \Idef{org129}Wayne State University, Detroit, MI, United States
\item \Idef{org130}Wigner Research Centre for Physics, Hungarian Academy of Sciences, Budapest, Hungary
\item \Idef{org131}Yale University, New Haven, CT, United States
\item \Idef{org132}Yonsei University, Seoul, South Korea
\item \Idef{org133}Zentrum f\"{u}r Technologietransfer und Telekommunikation (ZTT), Fachhochschule Worms, Worms, Germany
\end{Authlist}
\endgroup


\vspace{1cm}
\begin{thebibliography}{99}
\bibitem{GMVFNS}
B. A. Kniehl, G. Kramer, I. Schienbein and H. Spiesberger, 
Phys. Rev. Lett. 96 (2006) 012001;
B. A. Kniehl, G. Kramer, I. Schienbein and H. Spiesberger, 
Eur. Phys. J. C72 (2012) 2082.

\bibitem{fonll}
M. Cacciari, M. Greco and P. Nason, 
JHEP 05 (1998) 007; 
M. Cacciari and P. Nason,
JHEP 09 (2003) 006;
M. Cacciari, S. Frixione, N. Houdeau, M. L. Mangano, P. Nason and G. Ridolfi,
JHEP 10 (2012) 137.

\bibitem{Maciula:2013wg}
R. Maciula and A. Szczurek,
Phys. Rev. D 87 (2013) 094022.

\bibitem{Adams:2005dq} 
J.~Adams {\it et al.}  [STAR Collaboration],
Collaboration's critical assessment of the evidence from RHIC collisions,''
Nucl.\ Phys.\ A 757 (2005) 102;
B.~B.~Back {\it et al.} [PHOBOS Collaboration],
Nucl.\ Phys.\ A 757 (2005) 28;
K.~Adcox {\it et al.}  [PHENIX Collaboration],
Nucl.\ Phys.\ A 757 (2005) 184;
I.~Arsene {\it et al.}  [BRAHMS Collaboration],
Nucl.\ Phys.\ A 757 (2005) 1.

\bibitem{Aamodt:2010pa}
K.~Aamodt {\it et al.}  [ALICE Collaboration],
Phys.\ Rev.\ Lett.\ 105 (2010) 252302;
K.~Aamodt {\it et al.}  [ALICE Collaboration],
Phys.\ Lett.\ B 696 (2011) 30;
G.~Aad {\it et al.}  [ATLAS Collaboration],
Phys.\ Rev.\ Lett.\ 105 (2010) 252303.

\bibitem{aliceDRAA}
B. Abelev {\it et al.}  [ALICE Collaboration],
JHEP 09 (2012) 112.

\bibitem{Uphoff:2012gb} 
J.~Uphoff, O.~Fochler, Z.~Xu and C.~Greiner,
Phys.\ Lett.\ B 717 (2012) 430.

\bibitem{whdg}
S.~Wicks, W.~A.~Horowitz, M.~Djordjevic, and M.~Gyulassy, Nucl. Phys. A 784 (2007) 426; \\% nucl-th/0512076
W.~A.~Horowitz, AIP Conf. Proc. 1441 (2012) 889.

\bibitem{vitev}
R.~Sharma, I.~Vitev and B.~-W.~Zhang,
Phys.\ Rev.\ C 80 (2009) 054902.

\bibitem{He:2014cla}
M.~He, R.~J.~Fries and R.~Rapp,
arXiv:1401.3817.

\bibitem{ARNEODO}
M.~Arneodo,
Phys.\ Rept.\  240 (1994) 301.

\bibitem{Malace}
S. Malace, D. Gaskell, D. W. Higinbotham, I. Cloet,
arXiv:1405.1270.

\bibitem{EPS09} 
K. Eskola, H. Paukkunen and C. Salgado,
JHEP 04 (2009) 065.

\bibitem{deFlorianDS}
D. de Florian and R. Sassot,
Phys.Rev. D 69 (2004) 074028. 

\bibitem{HiraiHKN}
M. Hirai, S. Kumano and T.H. Nagai,
Phys.Rev. C 76 (2007) 065207. 

\bibitem{CGC2} 
H. Fujii and K. Watanabe, 
arXiv:1308.1258.

\bibitem{Tribedy}
P. Tribedy and R. Venugopalan, 
Phys. Lett. B 710, 125 (2012).

\bibitem{Albacete}
J. L. Albacete, A. Dumitru, H. Fujii, and Y. Nara, 
Nucl. Phys. A897, 1 (2013).

\bibitem{Rezaeian}
A. H. Rezaeian, Phys. Lett. B 718, 1058 (2013).

\bibitem{Vitev:2007ve}
I.~Vitev,
Phys. Rev. C 75 (2007) 064906.
  
\bibitem{Lev:1983hh}
M.~Lev and B.~Petersson,
Z.\ Phys.\ C 21 (1983) 155.

\bibitem{Wang:1998ww}
X.~N.~Wang,
Phys.\ Rev.\ C 61 (2000) 064910.

\bibitem{Kopeliovich:2002yh}
B.~Z.~Kopeliovich, J.~Nemchik, A.~Schafer and A.~V.~Tarasov,
Phys.\ Rev.\ Lett.\  88 (2002) 232303.

\bibitem{Arleo}
F. Arleo, S. Peigne, T. Sami,
Phys. Rev. D 83 (2011) 114036.

\bibitem{CMS:2012lrc}
S. Chatrchyan {\it et al.} [CMS Collaboration], 
Phys. Lett. B 718 (2013) 795.

\bibitem{Abelev:2012ola}
B. Abelev {\it et al.}  [ALICE Collaboration],
Phys. Lett. B 719 (2013) 29.

\bibitem{ABELEV:2013wsa}
B. Abelev {\it et al.}  [ALICE Collaboration],
Phys. Lett. B 726 (2013) 164.

\bibitem{Aad:2012gla}
G. Aad {\it et al.} [ATLAS Collaboration],
Phys.~Rev.~Lett. 110 (2012) 182302.
 
\bibitem{Adler:2006xd}
S. Adler {\it et al.}  [PHENIX Collaboration],
Phys. Rev. C 74 (2006) 024904.
 
\bibitem{Adare:2013ezl}
A. Adare {\it et al.}  [PHENIX Collaboration],
Phys. Rev. Lett. 111 (2013) 202301.

%\cite{Abelev:2014zpa}
\bibitem{Abelev:2014zpa} 
B.~B.~Abelev {\it et al.}  [ALICE Collaboration],
%``Suppression of $\psi$(2S) production in p-Pb collisions at $\sqrt{s_{NN}}$ = 5.02 TeV,''
arXiv:1405.3796 [nucl-ex].
%%CITATION = ARXIV:1405.3796;%%
%2 citations counted in INSPIRE as of 08 Aug 2014

\bibitem{phenixdAu}
A. Adare {\it et al.} [PHENIX Collaboration], 
Phys. Rev. Lett. 109 (2012) 242301.

\bibitem{phenixdAumuons}
A. Adare {\it et al.} [PHENIX Collaboration], 
arXiv:1310.1005.

\bibitem{STARDmesons_dAu}
J. Adams {\it et al.} [STAR Collaboration], 
Phys. Rev. Lett. 94 (2005) 062301.

\bibitem{AnneSickles}
A. M. Sickles, 
Phys. Lett. B 731 (2014) 51. 
%arXiv:1309.6924.

\bibitem{aliceJINST}
K. Aamodt {\it et al.} [ALICE Collaboration], 
JINST 3 (2008) S08002. 

\bibitem{alicePerformance}
B. Abelev {\it et al.} [ALICE Collaboration],
arXiv:1402.4476.

\bibitem{PDG}
J.~Beringer {\it et al.}, [Particle Data Group], Phys.\ Rev.\ D 86 (2012) 010001.

\bibitem{aliceDpp276}
B. Abelev {\it et al.} [ALICE Collaboration], 
JHEP 07 (2012) 191.

\bibitem{aliceDpp7}
B. Abelev {\it et al.}  [ALICE Collaboration],
JHEP 01 (2012) 128.

\bibitem{aliceDs7}
B. Abelev {\it et al.}  [ALICE Collaboration],
Phys. Lett. B 718 (2012) 279.

\bibitem{d'Enterria:2003qs}
  D.~G.~d'Enterria,
 arXiv:nucl-ex/0302016.

\bibitem{vdm}
B. Abelev {\it et al.}  [ALICE Collaboration], 
arXiv:1405.1849.

\bibitem{hijing} 
X.N. Wang and M. Gyulassy, 
Phys. Rev. D 44 (1991) 3501.

\bibitem {pythia} 
T. Sjostrand, S. Mrenna and P.Z. Skands, 
JHEP 05 (2006) 026.

\bibitem{Skands:2010ak}
P.~Z.~Skands,
Phys.\ Rev.\ D {\bf 82} (2010) 074018.

\bibitem {geant} 
R. Brun {\it et al.}, 
CERN Program Library Long Write-up W5013 (1994).

\bibitem{evtgen} 
D. Lange, Nucl. Instrum. Meth. A462 (2001) 152.

\bibitem{scaling}
R. Averbeck, N. Bastid, Z. Conesa del Valle, P. Crochet, A. Dainese and X. Zhang, 
arXiv:1107.3243.

\bibitem{MNR} 
M. Mangano, P. Nason and G. Ridolfi, Nucl. Phys. B 373 (1992) 295.

\bibitem{CTEQ6M}
D. Stump, J. Huston, J. Pumplin, W. K. Tung, H. L. Lai, S. Kuhlmann and J. F. Owens, 
JHEP 0310 (2003) 046.
\end{thebibliography}
\end{document}